\documentclass[runningheads]{llncs}
\usepackage{placeins}

\usepackage[T1]{fontenc}
\usepackage{algorithm,algorithmicx,algpseudocode}
\usepackage{amsmath}
\usepackage{caption} 
\usepackage{verbatim} 
\usepackage{subcaption}
\usepackage{graphicx}
\begin{document}
\title{Efficient Parallel Implementation of the Pilot Assignment Problem in Massive MIMO Systems}
\titlerunning{Parallel Pilot Assignment in Massive MIMO Systems}
\author{Eman Alqudah  \and
Ashfaq Khokhar
}

\authorrunning{E. Alqudah and A. Khokhar}

\institute{
  Department of Electrical and Computer Engineering, Iowa State University,\\
  Ames, IA-50011, USA \\
  \email{\{alqudah, ashfaq\}@iastate.edu}
}
\maketitle              % typeset the header of the contribution
\begin{abstract}
The assignment of the pilot sequence is a critical challenge in massive MIMO systems, as sharing the same pilot sequence among multiple users causes interference, which degrades the accuracy of the channel estimation. This problem, equivalent to the NP-hard graph coloring problem, directly impacts real-time applications such as autonomous driving and industrial IoT, where minimizing channel estimation time is crucial. This paper proposes an optimized hybrid K-means clustering and Genetic Algorithm (SK-means GA) to improve the pilot assignment efficiency, achieving a 29.3\% reduction in convergence time (82s vs. 116s for conventional GA). A parallel implementation (PK-means GA) is developed on an FPGA using Vivado High-Level Synthesis Tools (HLST) to further enhance the run-time performance, accelerating convergence to 3.5 milliseconds. Within Vivado implementation, different optimization techniques such as loop unrolling, pipelining, and function inlining are applied to realize the reported speedup. This significant improvement of PK-means GA in execution speed makes it highly suitable for low-latency real-time wireless networks (6G). 

\keywords{MIMO\and Pilot sequences \and Pilot assignment\and  Low-latency \and Real time \and Genetic algorithm \and K-means clustering algorithm \and FPGA \and high-level synthesis \and 5/6G Networks.}
\end{abstract}
\section{Introduction}

Fifth and sixth generation (5G/6G) wireless networks are designed to support ultra-high data rates and low-latency communication (URLLC)\cite{b32,b33} , vital for real-time applications such as autonomous driving, remote surgery, and industrial IoT \cite{b1}. A crucial technology enabling these requirements is massive multiple-input multiple-output (M-MIMO), which utilizes hundreds or even thousands of antennas at the base station (BS) to significantly improve throughput, spectral efficiency, and energy performance \cite{b3}.
M-MIMO allows concurrent data transmissions via spatial multiplexing and beamforming, improving user capacity without needing extra spectrum. As antenna counts grow, beam patterns become narrower and more targeted, reducing interference and latency \cite{b4}. However, such gains depend on accurate Channel State Information (CSI), which reflects the condition of the wireless channel \cite{b5}. In TDD systems, CSI is estimated via uplink pilot sequences, exploiting channel reciprocity to enable effective downlink transmission \cite{b5}.
One of the main challenges in CSI estimation is pilot contamination (PC), which arises when pilot sequences are reused across neighboring cells due to limited pilot resources. This leads to inter-cell interference, degrading CSI quality and system performance \cite{b3}. PC remains a bottleneck, even with an infinite number of antennas \cite{b8}.

For real-time systems, fast and accurate CSI estimation becomes increasingly difficult as user and antenna counts scale up \cite{b10}. Efficient pilot assignment can significantly mitigate PC, improve CSI accuracy, reduce pilot overhead, and enhance system responsiveness \cite{b10}. This is especially important for low-latency applications in emerging 5G/6G and IoT networks.
Pilot contamination mitigation in massive MIMO systems generally falls into three categories: channel estimation \cite{b11}, precoding \cite{b12}, and pilot scheduling \cite{b13}. This work focuses on pilot scheduling, where users in adjacent cells are reassigned pilot sequences to reduce interference and enhance spectral efficiency.

This paper makes the following key contributions:
\begin{enumerate}
\item \textbf{Real-Time Pilot Assignment Optimization:}  
We propose a 2D genetic algorithm with an elitism strategy to ensure optimal candidate solutions are preserved, reducing convergence time and enhancing robustness for real-time applications.

\item \textbf{Accelerated Convergence via K-means Integration:}  
K-means clustering is used to initialize the population of the genetic algorithm, improve the local data distribution, and accelerate convergence while lowering the computational load.

\item \textbf{FPGA-Based Low-Latency Implementation:}  
The proposed scheme is implemented using Vivado HLS optimizations. Different code modules were investigated and identified within the sequential implementation to apply relevant optimizations such as loop unrolling, pipelining, and function inlining to enable high-throughput and low-latency execution suitable for 5G/6G and IoT environments.

\item \textbf{Performance Gains and Complexity Reduction:}  
Simulation results show that our k-means-enhanced GA (KGA) achieves superior performance with a 29\% reduction in convergence time compared to existing methods, making it well-suited for real-time massive MIMO systems.
\end{enumerate}

The remainder of this paper is organized as follows. Section 2 reviews related work. Section 3 presents the pilot assignment problem, system model, and spectral efficiency formulation. Section 4 details the proposed KGA-based pilot allocation scheme. Section 5 discusses the FPGA implementation. Section 6 presents simulation results, and Section 7 concludes the paper.

\section{Related Work} 
Research on real-time communication Several approaches have been explored in \cite{b14,b15,b16,b17} to alleviate pilot contamination in TDD massive MIMO systems. In \cite{b14}, a time-shifted pilot assignment method was introduced, where pilot signals were transmitted asynchronously across neighboring cells. However, this approach could lead to interference between data and pilot signals. H. Yin \cite{b15,b16} proposed a covariance-based coordinated pilot assignment scheme, demonstrating that users with non-overlapping angles of arrival (AOAs) would not interfere with each other. Nevertheless, this method assumes a small AOA spread for each user.  

A graph coloring-based pilot allocation (GC-PA) scheme was examined in \cite{b17} to mitigate pilot contamination by efficiently assigning pilots among users within the interference graph. In \cite{b18}, a graph coloring-based scheme was proposed, where orthogonal pilot sequences were assigned to clusters of devices rather than individual devices. This allowed multiple devices to share the same pilot for periodic data transmission, leveraging the max k-cut graph partitioning technique to reduce pilot contamination in a multicell massive MIMO system. While this method enhanced spectral efficiency and improved scalability, it also increased system complexity.  

A spatial orthogonality-based greedy pilot allocation algorithm was proposed in \cite{b19}, using statistical channel covariance to maximize sum rate and reduce pilot contamination, though with high complexity. In \cite{b20}, three pilot scheduling strategies—greedy (GA), Tabu search (TSA), and greedy Tabu search (GTSA)—were studied, and a spatial orthogonality-based GA was proposed to balance complexity and performance.

In \cite{b21}, a single-dimensional genetic algorithm improved pilot assignment over random, greedy, and TS methods but required many iterations due to random initialization and lack of elitism. A bi-dimensional GA (2D-GA) in \cite{b22} incorporated TS to improve initial population convergence, at the cost of increased system complexity.
To address this, we propose a low-complexity, machine learning-based pilot allocation scheme using a two-dimensional parallel genetic algorithm (2D-PGA). The algorithm aims to maximize the minimum asymptotic signal-to-interference ratio (SIR) by optimizing the pilot assignment. To improve convergence, we integrate a k-means clustering method that groups the initial population into sub-populations, enhancing search efficiency. 
%The algorithm is implemented on an FPGA using Vivado High-Level Synthesis Tools (HLST) to enable real-time execution.

\section{Pilot Assignment Problem (PAP)} \label{sec:datamanagementoverview}

%\subsection{Channel State Information (CSI) and Pilot Signals}

\begin{figure}[htbp]
    \centering
    \includegraphics[width=\columnwidth]{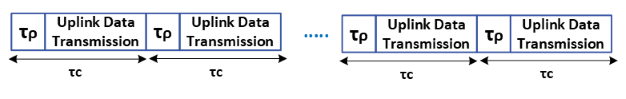}
    
    \caption{Coherence intervals in time-frequency space.}
    \label{fig:coherent}
\end{figure}

Massive MIMO systems require frequent estimation of the channel between each user and base station (BS), which is valid only over a short coherence time and bandwidth. In this regard, Channel State Information (CSI) is obtained by comparing the received pilot signals with known sequences. Each user must transmit a mutually orthogonal pilot sequence to avoid interference. However, because of limited pilot resources, orthogonal sequences are reused across cells, leading to pilot contamination. Uplink training is carried out during a portion \(\tau_\rho\) of the coherence interval \(\tau_c\), as illustrated in Fig.~\ref{fig:coherent}.
\vspace{6pt}

\subsection{System Model}

We consider a system with \(L\) cells, each with a BS of \(M\) antennas serving \(K\) single-antenna users. Each cell has \(K\) orthogonal pilot sequences \((\phi_1, \dots, \phi_K)\), reused across cells. Users are clustered, and each cluster is assigned distinct pilots reused in other cells. The channel vector between user \( k \) in cell \( j \) and base station (BS) \( i \) is defined as:

\begin{equation}
h_{ijk} = g_{ijk} \sqrt{\beta_{ijk}} \tag{1}
\end{equation}

where \(g_{ijk} \sim \mathcal{CN}(0, I_M)\) is small-scale fading and \(\beta_{ijk}\) is the large-scale fading factor:

\begin{equation}
\beta_{ijk} = \frac{z_{ijk}}{d_{ijk}^\alpha} \tag{2}
\end{equation}

\begin{figure}[htbp]
    \centering
    \includegraphics[width=0.95\columnwidth]{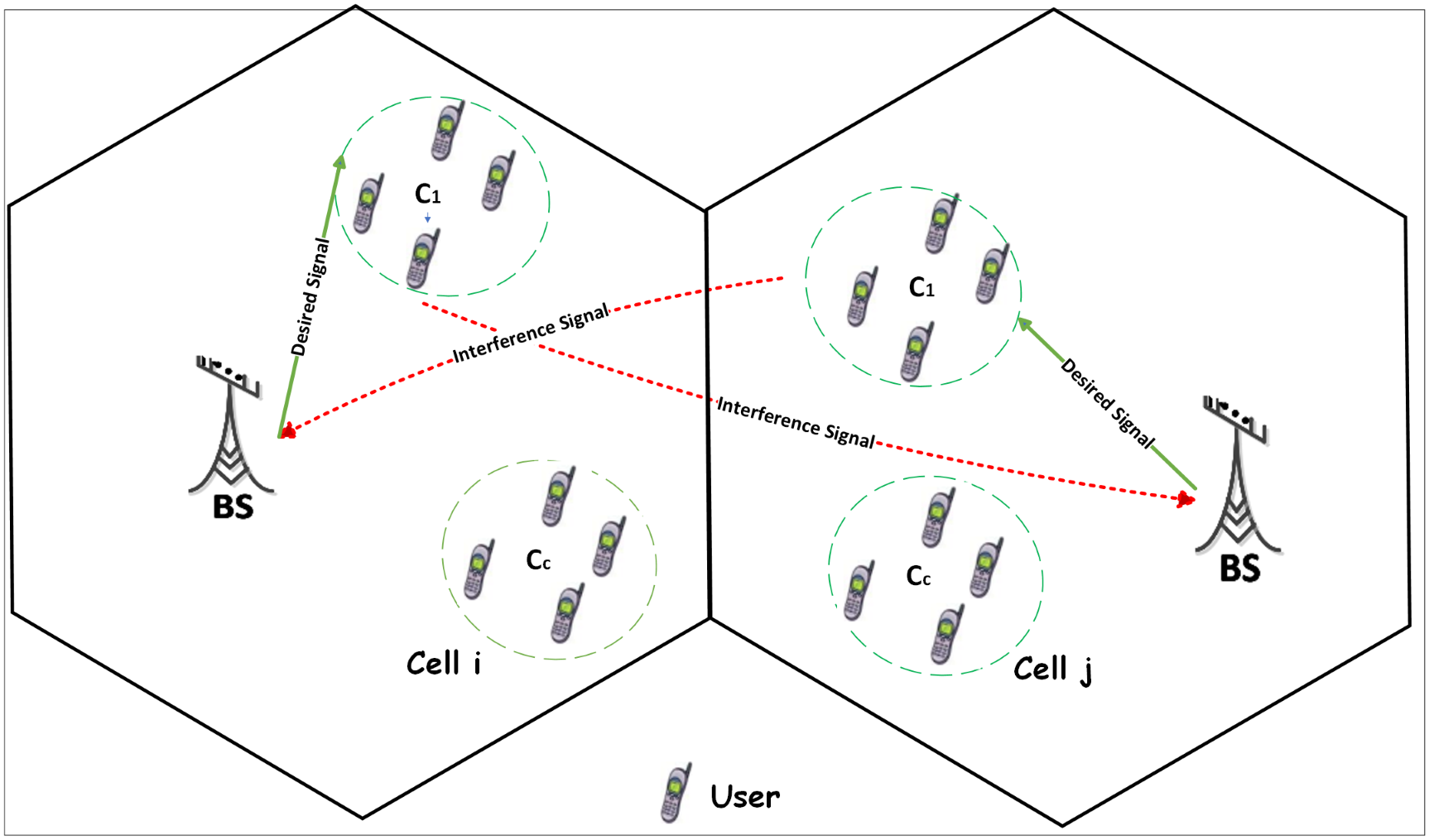}
    \caption{Clustered multi-cell, multi-user massive MIMO system.}
    \label{fig:fig2}
\end{figure}

Here, \(z_{ijk}\) models log-normal shadowing, \(d_{ijk}\) is user-BS distance, and \(\alpha\) is the path-loss exponent. Fig.~\ref{fig:fig2} shows the clustered multi-cell model. CSI is estimated via uplink pilots; however, reuse across cells introduces contamination.

\subsection{Uplink Spectral Efficiency}

As the number of BS antennas \(M\) grows, the user’s uplink SINR improves:

\begin{equation}
\text{SINR}_{jk}^{\text{up}} = \frac{|g_{jk}^H g_{jk}|^2}{\sum_{l \neq j} |g_{lk}^H g_{lk}|^2 + \sigma_{jk}^2} \tag{3}
\end{equation}

\begin{equation}
\text{SINR}_{jk}^{\text{up}} \to \frac{\beta_{jk}^2}{\sum_{l \neq j} \beta_{lk}^2} \quad \text{as } M \to \infty \tag{4}
\end{equation}

Thus, the uplink SE of user \(k\) in cell \(i\) is:

\begin{equation}
SE_{ik} = \log_2 \left( 1 + \frac{\beta_{iik}^2}{\sum_{j \neq i} \beta_{ijk}^2} \right) \tag{5}
\end{equation}

The total system sum-rate is:

\begin{equation}
SE_{\text{sum}} = \sum_{i=1}^{L} \sum_{k=1}^{K} SE_{ik} \tag{6}
\end{equation}

\subsection{Problem Formulation}

Each cell uses the same set of \(K\) orthogonal pilot sequences, uniquely allocated within a cell but reused between cells. The number of possible allocations is \(K!^{(L-1)}\) \cite{b17}, making exhaustive search infeasible in real time. Thus, the pilot assignment problem is formulated as maximizing the system sum-rate:

\begin{equation}
\max_{\{s_{ik}\} \in \{\phi_1, \dots, \phi_K\}} \sum_{i=1}^{L} \sum_{k=1}^{K} \log_2 \left( 1 + \frac{\beta_{iik}^2}{\sum_{j \neq i} \beta_{ijk}^2} \right) \tag{7}
\end{equation}

This formulation balances spectral efficiency and real-time constraints by optimizing pilot reuse to reduce contamination.

\section{Proposed Pilot Sequences Allocation Scheme}

Although exhaustive search methods guarantee optimal pilot sequence assignment, their high computational complexity makes them unsuitable for large-scale, real-time massive MIMO systems. To address this, we propose a hybrid optimization approach combining a 2D Genetic Algorithm (GA) with K-means clustering. This balances solution quality with computational efficiency, making it suitable for latency-sensitive environments.

\subsection{Sequential K-means GA and Chromosome Encoding}

The GA is a population-based optimization technique inspired by natural selection \cite{b24}. Each chromosome encodes the assignment of $K$ pilot sequences to $C$ user clusters as a 2D integer matrix. The optimization iterates over generations to improve pilot allocation by minimizing interference.

To avoid premature convergence and focus the search, we incorporate K-means clustering to partition users into $C$ spatial or interference-based clusters. GA operations—selection, crossover, and mutation—are then applied locally within each cluster, forming the basis of our Sequential K-means GA (SK-means GA).

\vspace{2pt}
\begin{algorithm}
\caption{Sequential K-means Based GA for Pilot Assignment}
\label{alg:kmeans_ga_pilot}
\begin{algorithmic}[1]
\State \textbf{Input:} Number of clusters $C$, Cells $L$, Pilots $K$, Coefficient matrix $\mathbf{\beta}$
\State Initialize population randomly.
\While{termination not met}
    \State Apply K-means clustering
    \State Evaluate fitness and compute interference using $\mathbf{\beta}$
    \For{$i = 1$ to $C$}
        \State Select and evolve cluster $i$ (selection, crossover, mutation)
        \State Evaluate cluster fitness and update local optima
    \EndFor
    \State Update global optimum and refresh population
\EndWhile
\State \Return Best pilot assignment
\end{algorithmic}
\end{algorithm}
\FloatBarrier

\subsection{Fitness Evaluation Function}

The fitness function evaluates each chromosome by measuring inter-cluster interference using large-scale fading coefficients $\beta$, which incorporate both path loss and shadowing effects. The goal is to minimize interference between user clusters sharing the same pilot:

\begin{equation}
\min I_{(c, c')} = \sum_{c=1}^{C} \sum_{c'=c}^{C} \left| \frac{\sum_{k \in C_{c'}} \frac{\beta_{jk}^{i}}{|C_{c'}|}}{\sum_{k \in C_c} \frac{\beta_{ik}^{i}}{|C_c|}} \right| + \left| \frac{\sum_{k \in C_c} \frac{\beta_{jk}^{j}}{|C_c|}}{\sum_{k \in C_{c'}} \frac{\beta_{jk}^{j}}{|C_{c'}|}} \right| \tag{8}
\end{equation}

\subsection{Parallel K-means Clustered GA (PK-means GA)}

To further accelerate the algorithm, we introduce the Parallel K-means Clustered GA (PK-means GA), which partitions the population using K-means clustering and processes each cluster independently. The key difference from SK-means GA is that cluster-based GA evolution is performed in parallel. The pseudocode is shown in Algorithm~\ref{alg:pkmean}. Implementation details and parallel FPGA execution are described in Section~\ref{sec:realtime}.

\begin{algorithm}
\caption{Parallel K-means Based GA for Pilot Assignment}
\label{alg:pkmean}
\begin{algorithmic}[1]
\State \textbf{Input:} Clusters $C$, Cells $L$, Pilots $K$, Matrix $\mathbf{\beta}$
\State Initialize population
\While{termination not met}
    \State Apply K-means and compute fitness
    \For{each cluster $c = 1$ to $C$ \textbf{in parallel}}
        \State Apply GA operations: selection, crossover, mutation
        \State Evaluate fitness and local optimum
    \EndFor
    \State Update global optimum and population
\EndWhile
\State \Return Best assignment
\end{algorithmic}
\end{algorithm}

\section{Real-Time Parallel Integrated Pilot Allocation Scheme}
\label{sec:realtime}

This section describes the FPGA-based real-time implementation of the proposed pilot allocation scheme using the Vivado High-Level Synthesis (HLS) tool. The design is optimized for low latency and efficient interference management in dynamic environments.

FPGAs are widely adopted in real-time systems for their ability to support highly parallel computations using reconfigurable resources such as LUTs, FFs, DSPs, and RAMs \cite{b29}. Their low-latency nature, combined with reconfigurability and low Non-Recurring Engineering (NRE) cost, makes them ideal for adaptive and scalable wireless systems.

Modern HLS tools, such as Vivado HLS, allow hardware designers to implement complex algorithms in C/C++, which are then synthesized into RTL. Optimizations such as loop pipelining, dataflow, unrolling, and inlining are supported to improve timing and resource efficiency \cite{b30,b31}. Vendor tools also automate placement, routing, and timing analysis to accelerate development.

The proposed PK-means GA is particularly well-suited for hardware implementation. Each user cluster identified by K-means can be assigned to a dedicated processing unit on the FPGA, allowing each GA instance to evolve in parallel. The most computationally intensive part—fitness evaluation—is parallelized using pipelined architectures.

This parallelism enables real-time operation, especially in dynamic wireless environments requiring frequent pilot reassignment. Further implementation details and flowcharts are provided in Appendix details.

\section{Results and Discussion}

\subsection{Simulation Setup}

Simulations were conducted using MATLAB R2022a in a realistic multi-cell massive MIMO environment consisting of $L=16$ hexagonal cells. Each cell contains up to $K=60$ single-antenna users, with base stations (BS) equipped with up to $M=256$ antennas. The channel is modeled as i.i.d. Rayleigh fading with a pilot reuse factor of 1. The objective is to minimize total inter-cell interference caused by pilot contamination.

The baseline Genetic Algorithm (GA) applies real-valued encoding, roulette-wheel selection, single-point crossover ($p_c = 0.9$), and Gaussian mutation ($p_m = 0.02$) over $T=20$ generations with a population size of $N=120$. 

The SK-means GA integrates K-means clustering ($C=5$) every three generations to enhance convergence by guiding the search space. To enable real-time optimization, a parallelized version—PK-means GA—is implemented on a Xilinx Virtex-7 XC7VX690T FPGA using Vivado HLS 2020.2. The design operates at 200 MHz with 16-bit fixed-point arithmetic and 8 parallel pipelines for fitness evaluation and genetic operations.

\begin{table}[htbp]
\centering
\caption{System Simulation Parameters}
\label{tab:system_simulation_params}
\footnotesize
\begin{tabular}{|l|l|}
\hline
\textbf{Parameter} & \textbf{Value} \\
\hline
Number of cells, $L$ & 16 \\
Number of BS antennas per cell, $M$ & 64\textbf{--}256 \\
Users per cell, $K$ & 1\textbf{--}60 \\
Number of clusters, $C$ & 5 \\
Population size of GA, $N$ & 120 \\
Number of iterations of GA, $T$ & 20 \\
Mutation probability, $p_m$ & 0.02 \\
Crossover probability, $p_c$ & 0.9 \\
\hline
\end{tabular}
\end{table}

\subsection{Performance Evaluation and Comparison}

We evaluate the proposed methods in terms of convergence speed, assignment quality, and scalability. The following schemes are considered:

\begin{itemize}
    \item \textbf{Random Pilot Assignment (RPA):} Baseline method assigning pilots arbitrarily.
    \item \textbf{Exhaustive Pilot Assignment (EX-PA):} Brute-force approach; infeasible for large-scale $K$.
    \item \textbf{Traditional GA:} Heuristic search using evolutionary operators.
    \item \textbf{SK-means GA:} GA guided by clustering to accelerate convergence.
    \item \textbf{PK-means GA:} FPGA-accelerated implementation for real-time optimization.
\end{itemize}

The SK-means GA converges in 82 seconds, a 29.3\% improvement over the traditional GA (116 seconds). PK-means GA significantly accelerates convergence to just 3.5 milliseconds, demonstrating suitability for latency-sensitive systems. As the user load increases, EX-PA becomes computationally impractical, while the proposed methods maintain efficiency and solution quality.

\begin{figure}[htbp]
    \centering
    \includegraphics[width=0.75\columnwidth]{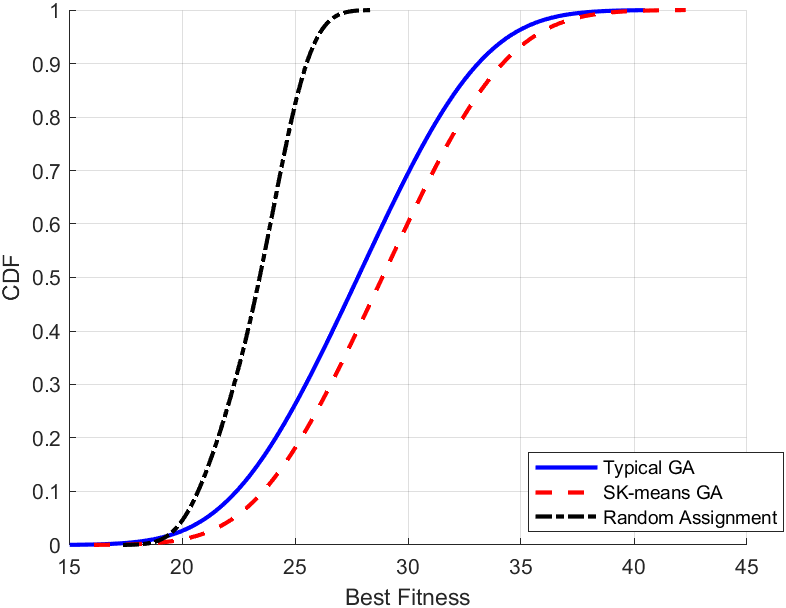}
    \caption{CDF of best fitness values for Random Assignment, Traditional GA, and SK-means GA.}
    \label{fig:fitness_cdf}
\end{figure}

\subsection{Hardware Resource Utilization}

The FPGA implementation of PK-means GA demonstrates the feasibility of high-throughput parallelism for evaluating large populations. We envision that with the increase in the population size or the number of clusters, hardware complexity and resource utilization will increase propoationately. particularly for LUTs and DSPs. The detailed analysis of these aspects of the implementation will be presented in the full version of the paper. Table~\ref{tab:resource_utilization} summarizes resource consumption on the Virtex-7 platform for 5 clusters and 120 users.

\begin{table}[htbp]
\centering
\caption{FPGA hardware resource utilization for PK-means GA using Vivado HLS for $C=5$.}
\label{tab:resource_utilization}
\renewcommand{\arraystretch}{1.2}
\setlength{\tabcolsep}{6pt} % Adjust spacing as needed
\footnotesize
\begin{tabular}{|l|c|c|c|c|}
\hline
\textbf{Component Name} & \textbf{BRAM\_18K} & \textbf{DSP 48E} & \textbf{FF} & \textbf{LUT} \\
\hline
Expression      & --    & --    & 0        & 44,402   \\
FIFO            & --    & --    & --       & --       \\
Instance        & 429   & 358   & 107,937  & 127,062  \\
Memory          & 24    & --    & 14       & 1        \\
Multiplexer     & --    & --    & --       & 6,506    \\
Register        & 0     & --    & 21,340   & 640      \\
\hline
Total           & 453   & 358   & 129,291  & 178,611  \\
Available       & 2,060 & 2,800 & 607,200  & 303,600  \\
Utilization (\%)& 21    & 12    & 21       & 58       \\
\hline
\end{tabular}
\end{table}

Although direct power consumption measurements were not conducted in this study, it is important to consider the correlation between hardware resource utilization and energy efficiency. %As shown in Table~\ref{tab:resource_utilization}, logic usage (e.g., LUTs and DSPs) increases significantly with the use of parallel pipelines and clustering operations. 
Higher utilization typically leads to increased dynamic power consumption, particularly in the logic and interconnect fabric. Therefore, while PK-means GA achieves impressive acceleration, its energy efficiency must be evaluated in future work, especially in scenarios requiring ultra-low power operation. Incorporating power-aware design techniques,  such as optimized placement and routing or dynamic voltage scaling, may further optimize the hardware for green communication use cases.

\subsection{Computational Complexity Comparison}

We analyze the theoretical complexity of each pilot assignment method. The traditional GA has complexity \( \mathcal{O}(NTLK) \), where \( N \), \( T \), \( L \), and \( K \) denote population size, generations, number of cells, and users per cell, respectively. SK-means GA introduces a clustering overhead of \( \mathcal{O}(N C K d) \), where \( C \) is the number of clusters, and \( d \) is the feature dimension (e.g., coordinates). PK-means GA retains the same asymptotic complexity but divides computations across \( P \) parallel processing units.

Table~\ref{tab:complexity_comparison} summarizes the theoretical computational complexity of the evaluated pilot assignment algorithms. Here, \( P \) denotes the parallelism degree in the FPGA implementation. Although the number of operations remains the same asymptotically, parallelization reduces execution time dramatically.
While RPA has the lowest complexity, it performs poorly in interference-heavy environments. EX-PA guarantees optimal results but becomes infeasible as $K$ and $L$ grow. Both SK-means and PK-means GA offer scalable and interference-aware pilot assignments, with the FPGA implementation delivering millisecond-level response times even in large scenarios (e.g., $L=16$, $K=60$, $N=120$).
\begin{table}[htbp]
\centering
\caption{Computational Complexity of Pilot Assignment Algorithms}
\label{tab:complexity_comparison}
\footnotesize
\setlength{\tabcolsep}{6pt}
\begin{tabular}{|l|l|}
\hline
\textbf{Algorithm} & \textbf{Complexity} \\
\hline
Random Pilot Assignment (RPA) & \( \mathcal{O}(1) \) \\
Exhaustive Search (EX-PA) & \( \mathcal{O}(K!^{L-1}) \) \\
Typical Genetic Algorithm & \( \mathcal{O}(NTLK) \) \\
Serial K-means GA (SK-means GA) & \( \mathcal{O}(NTLK) + \mathcal{O}(NCKd) \) \\
Parallel K-means GA (PK-means GA) & \( \mathcal{O}\left(\frac{NTLK}{P}\right) + \mathcal{O}\left(\frac{NCKd}{P}\right) \) \\
\hline
\end{tabular}
\end{table}

\begin{figure}[htbp]
    \centering
    \includegraphics[width=0.75\columnwidth]{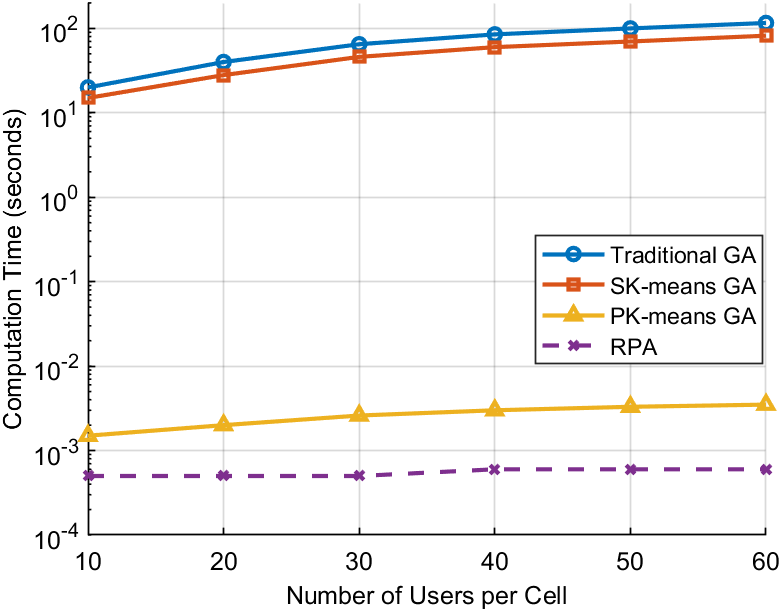}
    \caption{Computation time vs. number of users for all algorithms.}
    \label{fig:fig12}
\end{figure}

Figure~\ref{fig:fig12} illustrates the scalability trend: RPA remains constant but ineffective; EX-PA scales factorially; GA-based approaches scale linearly; and PK-means GA consistently offers the lowest runtime due to efficient parallel execution, making it suitable for 5G/6G systems with strict latency requirements.
\vspace{-0.95em}

\begin{figure}[htbp]
    \centering
    % Top row: Traditional GA and SK-means GA
    \begin{subfigure}[b]{0.45\columnwidth}
        \centering
        \includegraphics[height=5cm]{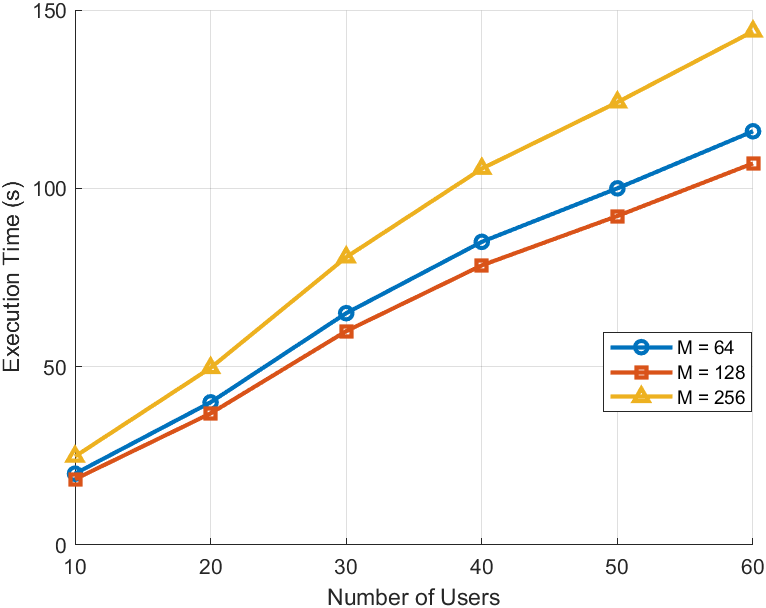} % set same height
        \caption{Traditional GA}
        \label{fig:fig13}
    \end{subfigure}
    \hfill
    \begin{subfigure}[b]{0.45\columnwidth}
        \centering
        \includegraphics[height=5cm]{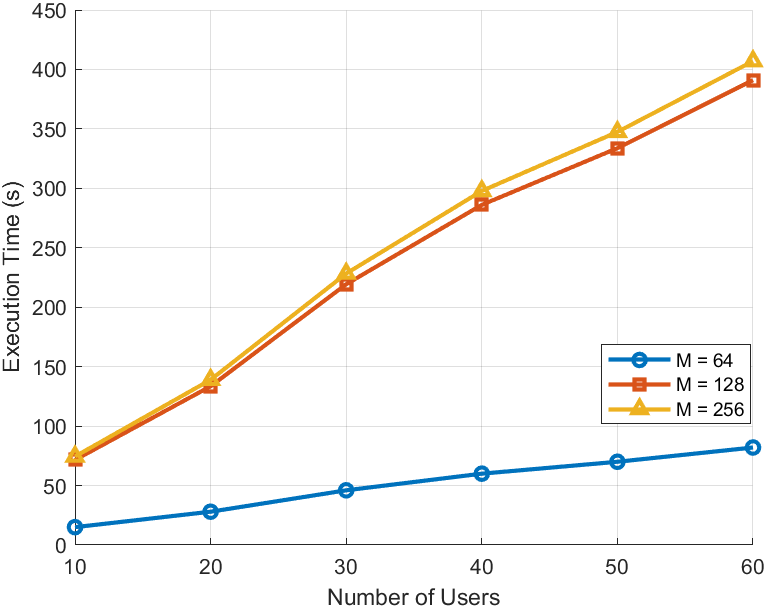} % same height
        \caption{SK-means GA}
        \label{fig:fig14}
    \end{subfigure}
    
    % Bottom row: PK-means GA
    \begin{subfigure}[b]{0.6\columnwidth}  % slightly wider and centered
        \centering
        \includegraphics[height=5cm]{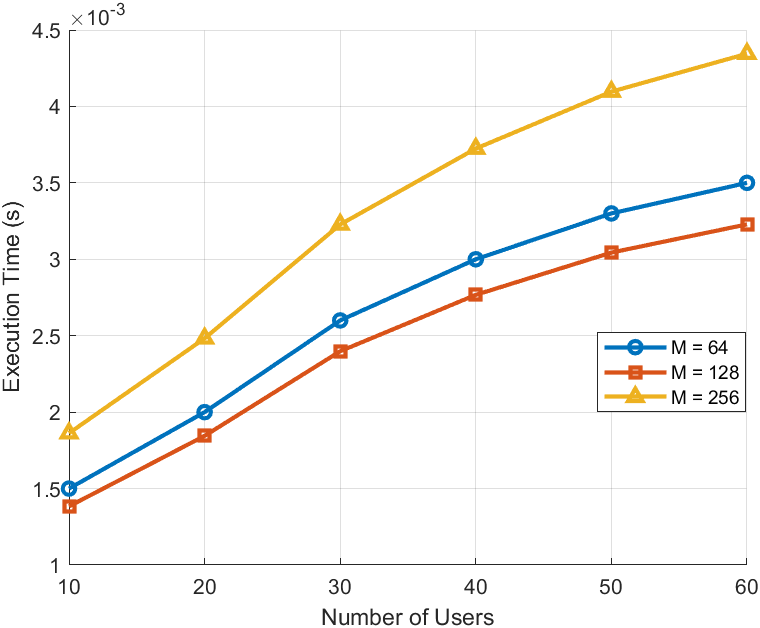} % same height for consistency
        \caption{PK-means GA}
        \label{fig:fig15}
    \end{subfigure}
    
    \caption{Computation time vs. number of users for different numbers of BS antennas $M$.}
    \label{fig:all_algorithms}
\end{figure}

Figure~\ref{fig:all_algorithms} presents the computation time trends for different GA-based algorithms across varying numbers of users and BS antennas. Traditional GA shows moderate execution times, while SK-means GA exhibits higher runtimes, especially for larger $M$. The PK-means GA consistently achieves the lowest execution times due to its parallelized implementation, demonstrating its scalability and suitability for real-time 5G/6G systems with stringent latency requirements.

\section{Conclusion}
This paper proposes an efficient parallel implementation of the pilot sequence assignment scheme by integrating the K-means clustering algorithm with the Genetic Algorithm (GA), termed PK-means GA, and optimizing it for real-time execution on FPGA. The algorithm was analyzed and streamlined to reduce computational overhead, while several optimization and parallelization techniques were employed to maximize speedup and efficiency. The Vivado HLS tool further accelerated the design process, providing optimization strategies to enhance performance.

Simulation results show that incorporating K-means into GA achieves a 29.3\% reduction in convergence time compared to traditional GA. The Cumulative Distribution Function (CDF) analysis confirms improved fitness performance, while the FPGA-based PK-means GA achieves convergence in only 3.5 milliseconds. This substantial runtime reduction, enabled by algorithmic optimizations and hardware-level parallelism, highlights the suitability of the proposed scheme for real-time deployment in MIMO base stations and time-sensitive 5G/6G applications.

%\subsection*{Limitations and Future Work}

%While the proposed PK-means GA framework performs effectively under standard simulation assumptions (e.g., hexagonal cells, i.i.d. Rayleigh fading), several areas remain open for improvement. Real-world networks may involve irregular topologies and coherent fading, which could affect performance. Moreover, the choice of $C=5$ clusters was empirically made without full sensitivity analysis, which is planned for future work.

%Scalability beyond $M=64$ antennas and $K=60$ users remains unexplored, and further optimization will be required for massive MIMO scenarios. Although our FPGA implementation demonstrates significant speedup, power consumption was not measured — an important metric for green communication systems.
%Future work will include power profiling and optimization to ensure that the proposed FPGA-based design aligns with green communication objectives.
%Finally, deep learning-based pilot assignment methods were not included in our benchmarking. Comparing such models — particularly in hardware-constrained environments — will be a valuable direction for future exploration.
\subsection*{Future Work}
This work emphasizes FPGA-based runtime acceleration, but does not report detailed power or energy efficiency results. Since these metrics are critical for 5G/6G systems, a comprehensive power analysis will be addressed in our extended work.

\begin{credits}
{\small\textbf{Disclosure of Interests.} The authors have no competing interests to declare that are relevant to the content of this article.}
\end{credits}

\end{document}